# Field emission from two-dimensional GeAs

Antonio Di Bartolomeo[1,2,*], Alessandro Grillo[1,2], Filippo Giubileo[2], Luca Camilli[3], Jianbo Sun[4], Daniele Capista [5] and Maurizio Passacantando [5,6]

[1] Department of Physics "E.R. Caianiello" and Interdepartmental Centre NanoMates, University of Salerno, via Giovanni Paolo II, Fisciano, 84084, Italy; Corresponding author: adibartolomeo@unisa.it

[2] CNR-SPIN, via Giovanni Paolo II, Fisciano, 84084, Italy

[3] Dipartimento di Fisica, Università degli studi di Roma "Tor Vergata", via della Ricerca Scientifica 1, Roma, 00133, Italy

[4] Department of Physics, Technical University of Denmark, Ørsteds Plads, 2800, Kgs. Lyngby, Denmark

[5] Department of Physical and Chemical Sciences, University of L'Aquila, via Vetoio, Coppito, L'Aquila, 67100, Italy

[6] CNR-SPIN L'Aquila, via Vetoio, Coppito, L'Aquila, 67100, Italy

* Author to whom correspondence should be addressed: adibartolomeo@unisa.it; Tel.: +39-089-96-9189

**Abstract**

GeAs is a layered material of the IV-V groups that is attracting growing attention for possible applications in electronic and optoelectronic devices. In this study, exfoliated multilayer GeAs nanoflakes are structurally characterized and used as the channel of back-gate field-effect transistors. It is shown that their gate-modulated p-type conduction is decreased by exposure to light or electron beam. Moreover, the observation of a field emission current demonstrates the suitability of GeAs nanoflakes as cold cathodes for electron emission and opens up new perspective applications of two-dimensional GeAs in vacuum electronics. Field emission occurs with a turn-on field of $\sim 80 \frac{V}{\mu m}$ and attains a current density higher than $10 \ A/cm^2$, following the general Fowler-Nordheim model with high reproducibility.

Germanium arsenide (GeAs) has emerged as an interesting layered compound of IV–V groups with a crystal structure belonging to the centrosymmetric monoclinic $C2/m(12)$ space group and strong in-plane anisotropy.[1,2] In GeAs, every Ge atom is coordinated to three As atoms and another Ge atom, while every As atom is coordinated to three Ge atoms (Fig. 1(a)). Each GeAs monolayer is terminated by As atoms and interacts with neighboring layers by weak van der Waals forces. The small interlayer cohesion energy (0.191 eV/atom) allows easy exfoliation. Ingots of GeAs have been synthesized by combining Ge and As in vacuum at high temperature (>1370 K)[2] and have been exfoliated in liquid phase or mechanically to obtain multilayer or few-layer nanosheets.[3] First-principles calculations and UV-visible absorption spectroscopy have demonstrated that GeAs nanosheets have a bandgap that increases significantly as the number of layers decreases, from 0.6 eV for the bulk up to 2.1 eV for the monolayer.[3,4] Remarkably, while the monolayer has a direct bandgap, the multilayers are predicted to have a quasi-direct bandgap.

The electrical properties of multilayer GeAs nanosheets have been measured to reveal their 2D carrier transport behavior[5] and anisotropic electrical conduction strongly affected by impurities.[6] Using multilayer GeAs field-effect transistors (FETs), it has been proved that the temperature-dependent conductivity can be described by the coexistence of variable range hopping among defect-induced bandgap states and band-like transport.[3,5,6] The carrier mobility in GeAs is higher along the zig-zag direction and typically in the range 0.1-10 $cm^2V^{-1}s^{-1}$.[5–7] The anisotropic crystal structure of GeAs leads also to highly anisotropic thermal conductivity,[2] mechanical response[8] and optical properties[7]. Moreover, stable and large photocurrents with rapid on/off switching as well as linear dichroism and polarized photodetection indicate that GeAs is a promising material for high-performance optoelectronic nanodevices, particularly for polarization optical applications.[3,9] GeAs nanosheets have also shown promising photoelectrochemical water splitting capability under visible light irradiation.[3,10]

In the present study, we characterize crystal structure and symmetry of mechanically exfoliated multilayer GeAs nanosheets. Field-effect transistors are obtained by contacting GeAs nanoflakes with Au electrodes and/or metallic tips. We investigate the field emission properties of GeAs nanoflakes, taking advantage from a nanotip inside a scanning electron microscope (SEM) chamber that is used as the anode for local field emission characterization.[11,12]

Field emission (FE) is the extraction of electrons from a semiconducting or metallic material under the application of an electric field. As a macroscopic manifestation of a quantum effect, FE offers significant scientific interests in material science and is exploited in many applications such as electron microscopy, electron spectroscopy, e-beam lithography as well as in vacuum electronics for nanoscale field emission transistors, displays and microwave generation or for x-ray tubes.[13–19] The externally applied electric field reduces the barrier for electron escape from the material to vacuum. FE is favored from electrically and thermally highly conducting materials with low work function and nanometer rough surface that give rise to local field enhancement. Hence, the intrinsic doping, the sharp edge and the low electron affinity below 4 eV of GeAs nanosheets are beneficial for field emission. Moreover, the GeAs electron affinity, which results in a low tunneling barrier, decreases with the number of layers, being only 3.17 eV for the bilayer and 2.78 eV for the monolayer.[3]

In this work, we show that a high and reproducible FE current can be extracted from the edge of GeAs nanoflakes.

Ultrathin GeAs flakes were mechanically exfoliated from bulk GeAs single crystals using a standard adhesive tape method.[20] The nanoflakes were exfoliated onto degenerately p-type doped silicon substrates, covered by 300 nm thick $SiO_2$. The $SiO_2$/Si substrate was endowed with patterned markers of 5 nm Ni/50 nm Au, which were previously defined by photolithography and lift-off. Sometimes, the transferred flakes ended up partially over the SiO$_2$ layer and partially over an Au marker establishing an electrical contact with it. The SEM images of two typical flakes are shown in Figs. 1(b) and 1(c). In particular, Fig. 1(b) displays a flake with minimal overlap with the Au marker. The contact with the Au marker is wider for the flake of Fig. 1(c), which has a uniform thickness of about 6 nm, corresponding to about 10 monolayers, as shown in Fig. 1(d) by its height profile as measured by atomic force microscopy (AFM).

The electrical conduction of the GeAs flakes was measured inside the SEM chamber at pressure below 10$^{-6}$ Torr and at room temperature. The flakes were gently contacted by means of two piezo-driven tungsten nanotips (Tip 1 and Tip 2) with nanometric movement control, used as the anode and the cathode, respectively. Such tips were connected to a semiconductor parameter analyser Keithley 4200-SCS, used as source-meter unit. To take advantage of the usually wider and more stable electrical contact produced by the van de Waals force between the GeAs nanosheet and the Au marker, we often chose flakes in electrical contact with a marker. In this configuration, the Au marker was used as the cathode and the tungsten tip in direct contact with the flake was used as the anode (inset of Fig. 1(b)). The Si substrate offered a third terminal and was used as the gate in a three-probe field-effect transistor configuration. A simple variation of this setup, with the tungsten tip (anode) detached from the flake and positioned at a fixed distance from the GeAs nanosheet (Fig. 1(c)) allowed the measurement of the local field emission current from the nanoflake.

The X-ray diffraction (XRD) was performed by means of a Bruker D5000 system equipped with Cu Kα (wavelength λ = 0.154 nm) line excitation source. The patterns were acquired in Bragg-Brentano mode that enables measurements of the crystal orientation of the GeAs flakes with respect to the substrate surface as well as of their lattice parameters. Raman spectroscopy measurements were performed using a LabRam high-resolution Micro Raman apparatus by Jobin Yvon with λ$_0$=632 nm excitation.

Fig. 1(e) reports the XRD spectrum of the sample that we used to determine the unit cell parameters by Rietveld refinement. The peaks identified in Fig. 1(e) correspond to the (-201) and (-402) base-centered monoclinic structure of GeAs, and the (111) planes of Au from the Au markers. The structure of GeAs was solved in the monoclinic space group C2/m (12) (JCPDS 011-0524) and the obtained lattice parameters are: a = 1.5552 nm, b = 0.3761 nm, c = 0.9524 nm and β =101.255. Furthermore, the XRD spectrum confirms an interlayer spacing of 0.66 nm in agreement with previously reported data.[20] These observations lead to the ball and stick model of Fig. 1(a), showing the GeAs crystal structure in the projection plane (010) along with the cell edges (red line). In Fig. 1(a), two different geometric orientations of the Ge-Ge bond can be clearly distinguished from the crystalline structure of the GeAs: one bond parallel and the other one perpendicular to the layer plane that is an evidence of the anisotropic nature of GeAs crystal structure. The Raman spectrum shown in Fig. 1(f) evidences multiple Raman active modes peaks due to highly asymmetric structure, with eight of them being A$_g$ modes (95, 106, 148, 175, 273, 276, 285 and 309 cm$^{-1}$), and one being B$_g$ mode (259 cm$^{-1}$). XRD and Raman measurements confirm that the GeAs flakes, exfoliated from crystal and transferred onto the substrate, have maintained intact their structural properties.

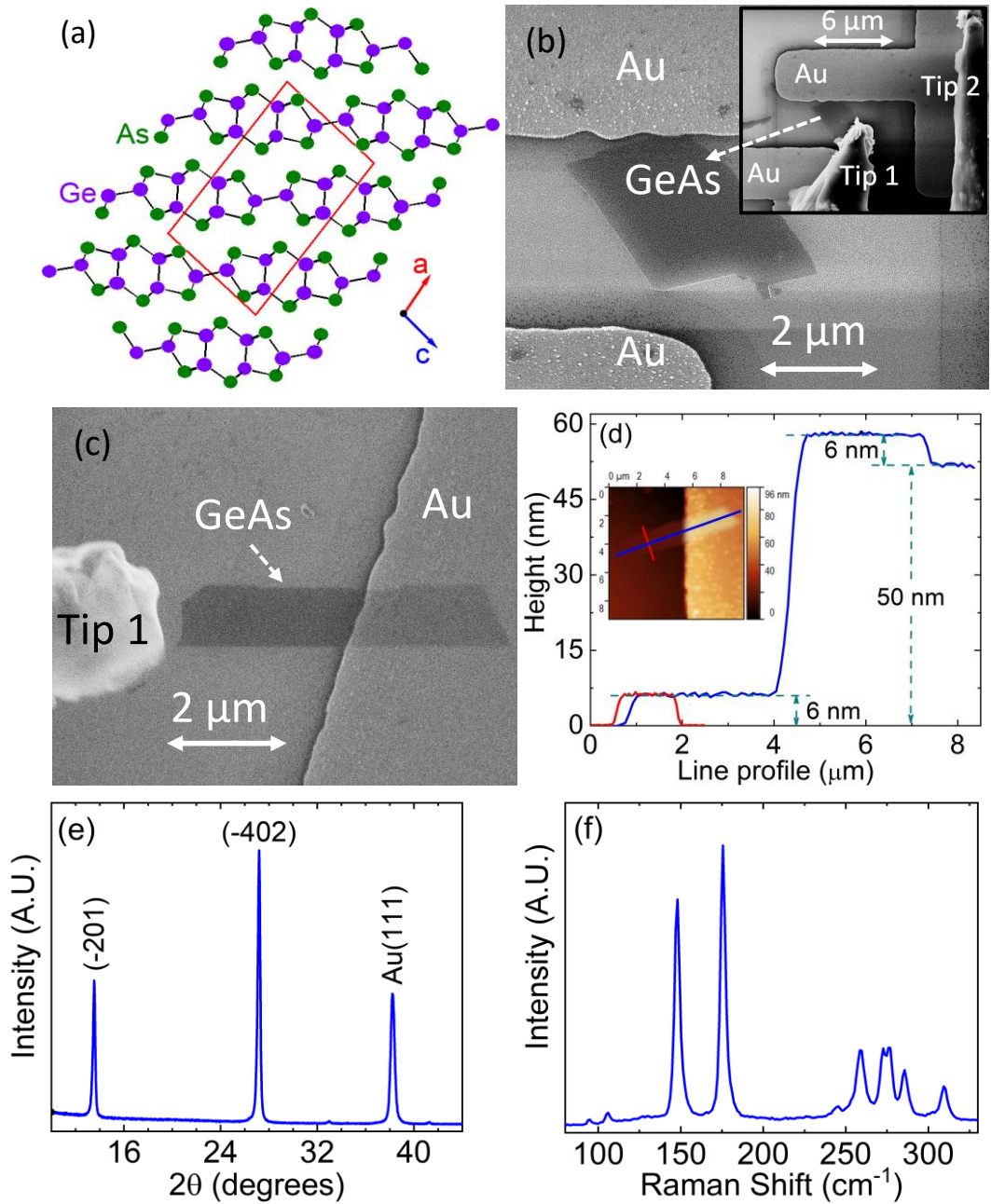

Fig. 1. (a) Side view for the crystal structure of layered GeAs. (a) SEM image showing a GeAs flake in tiny contact with the Au electrode (top marker). The inset shows the same flake contacted by a tungsten tip (Tip1); a second tungsten tip (Tip 2) is on the Au marker. The Au marker is used as the cathode while Tip 1 is used as the anode in the electrical measurements. (c) SEM image of another GeAs flake used for field emission measurements, with anode tip at hundreds nm from the flake. (d) AFM profile of the GeAs flake of Fig. (c), showing the nanosheet thickness of 6 nm. (e) XRD pattern in Bragg-Brentano geometry and (f) Raman spectrum of the same GeAs flake, measured in the parallel configuration.

The electrical conductance of the GeAs nanoflake of Fig. 1(a), used as the channel of a back-gate field-effect transistor, was measured in the common-source configuration schematically displayed in Fig. 2(a). The output characteristics, that is $I_{ds} - V_{ds}$ curves for fixed $V_{gs}$, of Fig. 2(b) show that the channel current is modulated by the gate and increases for negative $V_{gs}$, a behavior that is typical of a channel with hole conduction and has been seen already in similar devices.[5,6] The non-linearity of $I_{ds} - V_{ds}$ curves can be explained considering that the anode and cathode contacts are realized with different metals and have different areas. The $I_{ds} - V_{ds}$ asymmetric behavior is indicative of slightly different Schottky barriers formed at the W(tip)-GeAs and Au/GeAs interfaces.[21,22] The p-type conduction of the GeAs nanoflake is confirmed by Fig. 2(c), reporting the transfer characteristic, $I_{ds} - V_{gs}$ curve for a fixed $V_{ds}$, and showing that the transistor current decreases for increasing $V_{gs}$. The modulation of the current is limited to one order of magnitude, despite the quite large $V_{gs}$ range, consistently with the multilayer nature of the GeAs flake corresponding to a bandgap around several hundreds meV. From the slope of the transfer characteristic, we estimated the field-effect mobility as $\mu = \frac{L}{WC_{ox}V_{ds}} \frac{dI_{ds}}{dV_{gs}} \approx 0.6 \frac{cm^2}{Vs}$, where $C_{ox} = 11.5 \, nFcm^{-2}$ is the SiO$_2$ capacitance per unit area, $L \sim 2 \, \mu m$ and $W \sim 0.5 \, \mu m$ are the channel length and width, roughly measured from the SEM images. Such a value is at the low side of the range reported in the literature[3,5,7] being likely affected by the high contact resistance.[23]

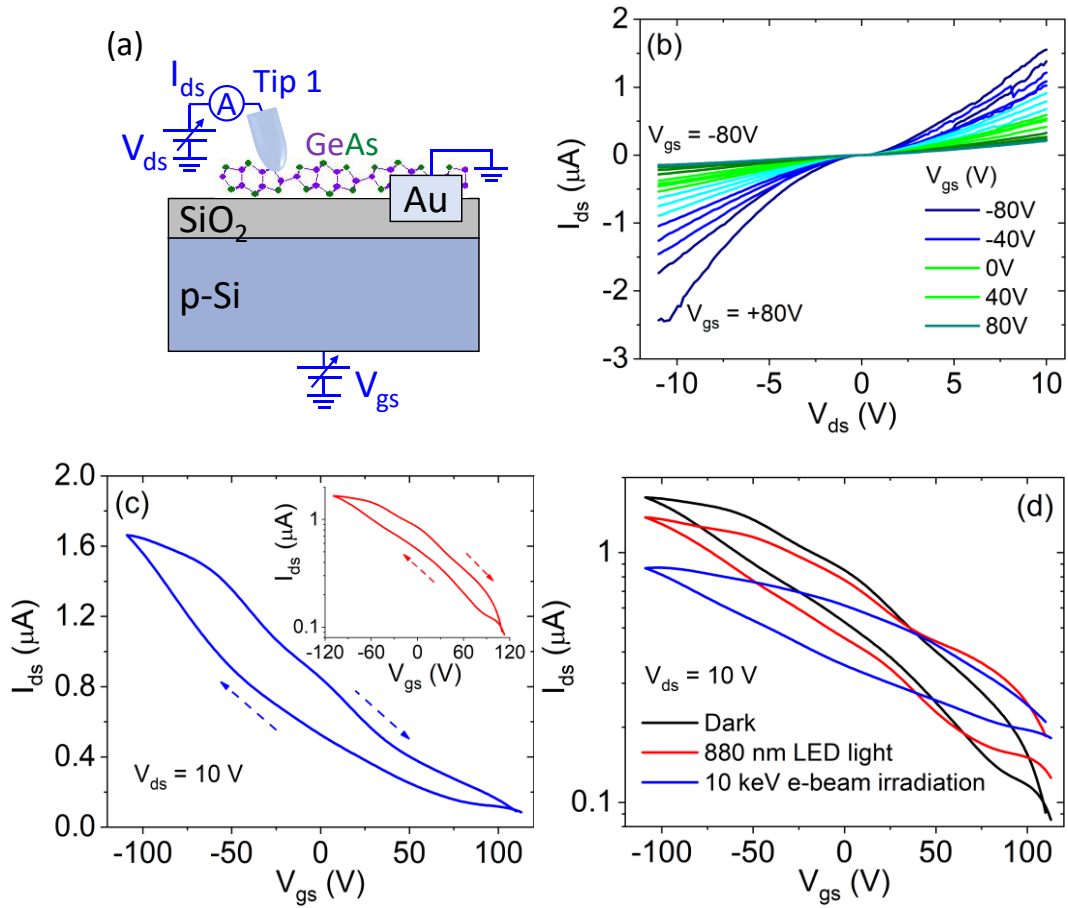

Fig. 2. (a) Schematic of the back-gate transistor used for the electrical characterization of the GeAs nanosheets. Electrical characterization of the nanoflake of Fig. 1(b): (b) Output characteristics $I_{ds} - V_{ds}$ for stepping $V_{gs}$, showing gate current modulation; (c) transfer characteristics $I_{ds} - V_{gs}$ for a fixed $V_{ds}$ with the current on linear and logarithmic scale (inset); (d) transfer characteristics in dark, under 880 nm LED light and after 10 keV electron beam (e-beam) irradiation. All measurements are performed at low pressure below $10^{-6}$ Torr and at room temperature.

Figure 2(c) shows clockwise hysteresis between the forward and reverse sweeps. Hysteresis in the transfer characteristics is a well-known feature of FETs with 2D-material channels and can be caused by gate-induced trapping/detrapping of free carriers in gap states.[24–27] Gap states are caused by external impurities and structural defects or are induced by surface adsorbates and interaction with the gate dielectric. However, electrical measurements were performed in a high vacuum, and adsorbates like water or oxygen should play a minor role while gap trap states from impurities[6] or GeAs/SiO$_2$ interaction should dominate.

Light[28] and electron beam irradiation[29–31] can strongly affect 2D-material transistors. Their impact on GeAs is investigated in Fig. 2(d). Both 880 nm LED light and 10 keV electron beam irradiation result in a reduction of the conductivity in the on state and a right shift of the transfer curve. This behavior corresponds to negative "photo"conductivity and can be easily explained by a gating effect.[32–34] The interaction of light or electron beam in the Si substrate generates electron-hole pairs. The reduced mobility of holes in Si and the vertical band bending give rise to a pile-up of holes at the SiO$_2$/Si interface and trapping in the SiO$_2$ dielectric. Such an accumulated charge acts as an extra gate with a positive voltage and eventually reduces the channel conductance.

The $I_{ds} - V_{ds}$ output curves shown in Fig. 3(a) were measured with Tip 1 in direct contact with the flake of Fig. 1(c) and Tip 2 on the Au marker, while the gate was grounded; the curves confirm the nonlinear p-type behavior of the GeAs nanoflakes as well as their negative photoconduction, as evidenced by the reduced current under 880 nm LED illumination. After the $I_{ds} - V_{ds}$ measurements, Tip 1 was detached from the flake and brought at about 400 nm from its edge (see Fig. 1(c) and inset of Fig. 3(b)). With the gate grounded, the voltage of Tip 1 (the anode) was slowly increased up to 110 V, while monitoring the current. Fig. 3(b) shows that the current remains at the noise floor of the experimental setup up to ~65 V, then it increases exponentially for more than 3 orders of magnitude. The behavior of the current is typical of field emission. The phenomenon is reproducible, with the FE current turning on at a slightly lower voltage (~52 V) for the successive anode voltage sweeps. The small reduction of the current with respect to the first sweep is likely due to an electrical conditioning effect, resulting in the desorption of residues or adsorbates at the edge of the flake by Joule heating during the first sweep.

Field emission is commonly described by the Fowler-Nordheim (FN) theory, which considers electron tunneling from a flat surface through a triangular barrier and relies on the Sommerfeld's free-electron theory for the description of the electronic distribution. Although emission from rough, irregular surfaces typical of modern nanoscale electron sources might require more complex equations,[35,36] the simplest FN theory[37] still provides a satisfying first-approximation model. According to it, the field emission current can be expressed as:

$$I = S \cdot A\phi^{-1} \left(\beta \frac{V}{kd}\right)^2 exp\left[-B\, \phi^{3/2} \left(\beta \frac{V}{kd}\right)^{-1}\right] \qquad (1)$$

where $S$ is the emitting area, $A = 1.54 \times 10^{-6}\, AV^{-2}eV$ and $B = 6.83 \times 10^9\, eV^{-3/2}m^{-1}V$ are dimensional constants, $\phi$ is the material work function, $V/kd = E$ is the electric field due to the applied voltage $V$ when the anode-cathode separation distance is $d$ with $k \sim 1.6$ a phenomenological factor accounting for the spherical shape of the anode.[38] β represents the so-called field enhancement factor due to the accumulation of the electrical field lines on the sharper protrusions of the emitting surface.[39] Such protrusions are the main emitting sites and in the present application correspond to the edge of the nanoflake. The FN eq. (1) well fits the experimental data, as shown by the red dashed curve in Fig. 3(b). The so-called FN plot of $\ln(IV^{-2})\ vs.\ V^{-1}$, shown in Fig. 3(c), further corroborates the FN nature of the observed current.

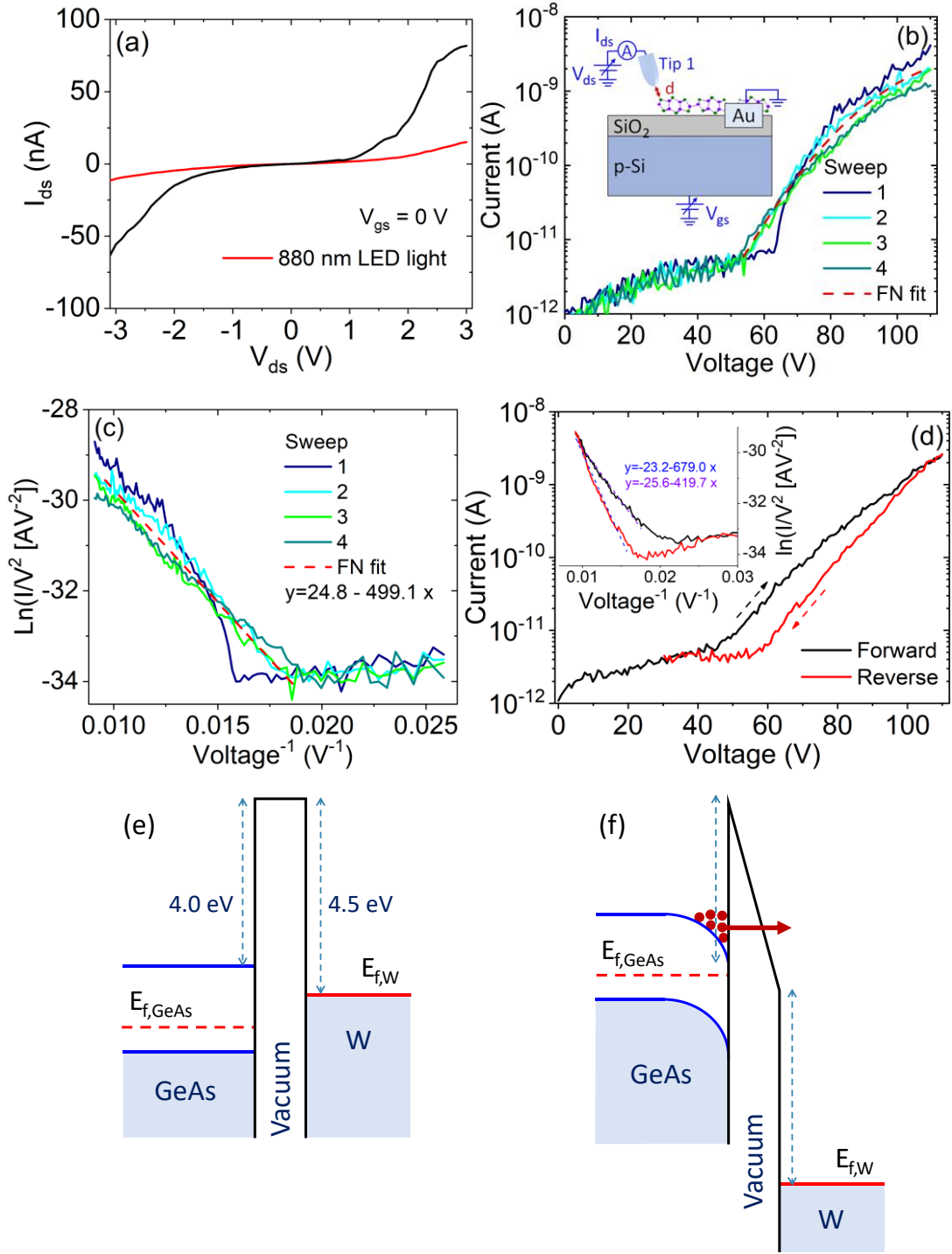

Fig. 3. Electrical characterization of the GeAs nanoflake of Fig. 1(c). (a) Output characteristics $I_{ds} - V_{ds}$ at $V_{gs} = 0$ V, in dark and under 880 nm LED illumination with Tip 1 in contact with the flake. (b) Current-Voltage measurements with Tip 1 at 400 nm from the edge of the flake showing a stable field emission current for V>50 V. (c) Fowler-Nordheim plot. (d) Current-Voltage measurements and Fowler-Nordheim plot (inset) for forward and backward Tip 1 voltage swept. Band diagram of the GeAs/vacuum/Tip 1 system without (e) and with (f) positive bias on Tip 1.

Fig. 3(d) displays the FE current and the FN plot (inset) over a forward and reverse sweep. Interestingly, the FE current decreases during the reverse sweep. This hysteretic behaviour could arise from space charge, which suppresses the emission during the ramp down of the anodic voltage.

The FN plot allows to estimate the field enhancement factor as $\beta = -Bkd\phi^{3/2}m^{-1}$, where $m$ is the slope of the fitting (dashed red) straight line. Assuming $\phi$=4.0 eV for multilayer GeAs, $\beta \sim 70$, according to the measurements of Fig. 3(c).

The turn-on field, defined as the voltage to which the current emerges from the setup noise floor, is $E_{to} \sim 80 \, V/\mu m$. Optimized emitters can have a turn-on field of few $V/\mu m$.[40,41] The p-type doping of GeAs contributes to increase $E_{to}$. Indeed, extraction of electrons from a p-type materials requires the achievement of an inversion condition at the emitting surface. Such an inversion occurs when the anode voltage is high enough to induce the required band bending, as shown in the energy band diagrams of Figs. 3(e) and 3(f). The inversion layer at the GeAs interface provides the electrons to tunnel through the vacuum barrier at high bias.

We finally note that both the $\beta$ factor and the turn-on field of GeAs nanoflakes are rather competitive when compared with those measured from other 2D materials in similar experimental conditions.[12,17,42–45] Furthermore, conservatively assuming that the emission occurs from the entire edge of the flake, the extracted current density attains the appreciable value of $\sim 10 \, A/cm^2$. The overall field emission figures of merit of GeAs result comparable to the field emission performance of carbon nanotubes,[46–49] Mo tips,[50] bare and metal coated Si tips,[51] or other established emitter materials.[52–57]

In conclusion, mechanically exfoliated multilayer GeAs nanosheets, transferred onto SiO$_2$/Si substrates, have been used as the channel of back gated transistors and as cold cathodes for electron emission. It has been found that the nanosheets possess intrinsic p-type doping and their conductivity is modulated by a gate. A reproducible field emission current from the edge of GeAs nanoflakes occurring with a turn-on field around $80 \, V/\mu m$ and attaining a remarkable current density higher than $10 \, A/cm^2$ has been reported.

This study provides experimental evidence of FE from GeAs nanoflakes and paves the way for new applications of 2D GeAs introducing it to the realm of vacuum electronics.

**Data availability**

The data that support the findings of this study are available from the corresponding author upon reasonable request.

**Acknowledgements**

This research was funded by the Italian Ministry of Education, University and Research (MIUR), projects Pico & Pro ARS01_01061 and RINASCIMENTO ARS01_01088. L.C. and J.S. want to express their gratitude to the Villum Fonden (Young Investigator Program, Project No. 19130). L.C. acknowledges support from MIUR via "Programma per Giovani Ricercatori - Rita Levi Montalcini 2017".